\documentstyle[aps,prl]{revtex}

\begin{document}
\twocolumn
\draft
\input{epsf}
\newcommand{\bt}{\mbox{\boldmath{$t$}}}
\newcommand{\bs}{\mbox{\boldmath{$s$}}}
\newcommand{\bn}{\mbox{\boldmath{$n$}}}
\newcommand{\bz}{\mbox{\boldmath{$z$}}}
\newcommand{\br}{\mbox{\boldmath{$r$}}}
\newcommand{\hs}{{\hat{s}}}

\twocolumn[\hsize\textwidth\columnwidth\hsize\csname@twocolumnfalse\endcsname

\title
{Error-Correcting Codes That Nearly Saturate Shannon's Bound} 

\author{Ido~Kanter$^{1}$ and David~Saad$^{2}$}
\address{
$^{1}$ Department of Physics, Bar-Ilan
University, Ramat-Gan 52900, Israel. \\ $^{2}$The Neural Computing Research
Group, Aston University, Birmingham B4 7ET, UK.}  

\maketitle

\begin{abstract}
Gallager-type error-correcting codes that nearly saturate Shannon's
bound are constructed using insight gained from mapping the problem
onto that of an Ising spin system. The performance of the suggested
codes is evaluated for different code rates in both finite and
infinite message length.

\end{abstract}
]

Efficient information transmission plays a central role in modern
society, taking a variety of forms, from telephone and satellite
communication to storing and retrieving information on disk-drives.
Error-correcting codes are commonly used in most methods of
information transmission to compensate for noise corrupting the data
during transmission; they require the use of additional information
transmitted together with the data itself. The percentage of
informative transmitted bits, determines the coding efficiency and
subsequentally the speed of communication channels and the effective
storage space on hard-disks.  In his seminal paper of 1948,
Shannon~\cite{Shannon} derived the channel capacity, providing bounds
on the code-rate for which codes, capable of achieving perfect
retrieval for a given noise level, can be found. The search for
efficient, practical error-correcting codes that saturate Shannon's
bound resulted in several practical codes, most of which are still
below Shannon's bound.  Here we propose a new approach based on
insight gained from the study of Ising spin-systems with
low-connectivity multi-spin interactions. Adapting our method to
Gallager's error-correcting codes~\cite{Gallager} one obtains codes
that nearly saturate the limits set by Shannon.

In a typical scenario, a message comprising $N$ binary bits is
transmitted through a noisy communication channel; the received string
differs from the transmitted one due to noise which may flip some
bits. We identify the flipping rate - $f \!\in\! \lbrack 0:1\rbrack$ -
in a binary symmetric channel as the fraction of bits that change
their value from 0 to 1 or from 1 to 0. We focus on this noise model
as it can be easily interpreted within the framework of Ising spin
systems; however, other noise types may also be considered, and may be
more realistic in some scenarios.  The receiver can correct the
flipped bits only if the source transmits $M(f)\!>\!N$ bits; the ratio
between the original number of bits and those of the transmitted
message $R\equiv N/M$ constitutes the code-rate for unbiased
messages. Shannon~\cite{Shannon} derived the channel capacity and
provided bounds on the maximal code rate $R_{c}$, for a given flip
rate $f$ and code bit error probability $p_b$, for which codes,
capable of achieving perfect retrieval, exist. The maximal code rate
equals the channel capacity and is given explicitly\cite{Cover} by
\begin{equation}
\label{eq:shannon_bound}
R_{c}=(1-H_2(f))/(1-H_2(p_b)) \ ,
\end{equation}
where $H_2(x)=x \log_{2}(x)+(1-x) \log_{2}(1-x)$.

Shannon's theory is unconstructive, and the many good algorithms that
have been introduced over the years (e.g., BCH, Reed-Muller and
Reed-Solomon codes, for a review see\cite{err_cor_book}) fall short of
saturating Shannon's bounds, although they may provide
close-to-optimal performance in specific scenarios. Even the most
advanced code to date, the Turbo code\cite{turbo} is somewhat below
Shannon's bound.

One error-correcting code which recently became popular is the
Gallager code~\cite{Gallager,MacKay,Davey,Shokrollahi}, which was
abandoned shortly after its introduction due to the limited
computational abilities of the time.  In this method, representing a
special case of parity-check codes, the transmitted message comprises
the original message itself and additional bits, each of which is
derived from the parity of a sum of certain message-vector bits.
The choice of the message-vector elements used for generating
single code-word bits is carried out according to a predetermined
random set-up and may be represented by a product of a randomly
generated sparse matrix and the message-vector in a manner explained
below. Decoding the received message relies on iterative probabilistic
methods like belief propagation\cite{MacKay,Frey}.

It has been shown that by using Gallager-type methods and specific
choices of the encoding/decoding matrix it is possible to improve the
maximal practically achievable code-rate~\cite{MacKay,Shokrollahi}
although results are still somewhat below Shannon's capacity.  The
root of the problem is the inevitable tradeoff between improving the
code's corrective capabilities and the need for a practical and
reliable iterative decoding process, guaranteed to converge from any
initial condition (i.e., that will not require additional, typically
unavailable, information about the message itself).  This goal is
achieved by understanding the physical characteristics of the problem
and devising a new method based on this insight. As Gallager-type
methods form the basis of our proposal we will now explain explicitly
the version we employ - the MN code~\cite{MacKay}.

In the MN code one constructs two sparse matrices $A$ and $B$
of dimensionalities $M\!\times\! N$ and $M\! \times\! M$
respectively. The matrix $A$ has $K$ non-zero (unit) elements per row
and $C (=KM/N)$ per column while $B$ has $L$ per row/column. The matrix
$B^{-1}A$ is then used for encoding the message
\[ {\bf t} = B^{-1}A \ \bs \ \mbox{(mod 2)} \ .  \]
The received message comprises the transmitted vector corrupted by the
noise vector $\bn$: $ {\bf r} = {\bf t} + {\bf n} \ \mbox{(mod 2)} \
.$
Decoding is carried out by employing the matrix $B$ to obtain:
$ {\bf z} = B  \ ({\bf t}+{\bf n}) = A \bs + B {\bf n} \ ,$
and requires solving the equation 
\[ [A , B] \left[ \begin{array}{c}
{\bs '} \\ \bn '
\end{array} \right] =  {\bf z}  \ , \]  
where $\bs '$ and $ \bn '$ are the unknowns. This may be carried out
using methods of belief network decoding\cite{MacKay,Frey}, where
pseudo-posterior probabilities, for the decoded message bits being
0 or 1, are calculated by solving iteratively a set of equations for
the conditional probabilities of the codeword bits given the decoded
message and vice versa. For exact details of the method used and the
equation themselves see\cite{MacKay}.

Most studies of Gallager-type codes have been carried out via methods of
information theory (e.g.,\cite{MacKay}). The first link between a
special case of Gallager's method, where $B=I$ the identity matrix,
and the realm of physical spin-systems was established by
Sourlas\cite{Sourlas} by mapping the problem onto that of a
Hamiltonian system, replacing the original Boolean variables by binary
ones which are analogous to spins in Ising-type systems with
Multi-Spin Interactions (MSI). For this simple case the system is
described by the Hamiltonian
\begin{equation}
\label{eq:hamiltonian}
H= -\sum_{\langle i_1,i_2 \ldots i_K \rangle}
J_{i_1,i_2,...,i_K} \ \hs_{i_1}' \hs_{i_2}' ...\hs_{i_K}'
\end{equation}
where $\{\hs_{i}' \}$ are the binary dynamical variables $(\pm 1)$,
used in the decoding process. The tensor $J_{i_1,i_2,...,i_K}= \pm
\hs_{i_1}\hs_{i_2} ...\hs_{i_K}$ with probabilities $1\!-\!f$ and $f$
correspondingly, represents the received codeword corrupted by noise
during transmission, ${\bf \hs}$ being the binary representation of
the original Boolean message vector $\bs$; the choice of indices
$i_1,i_2,...,i_K$ corresponds to the non-zero row elements of the
matrix $A$.  Under a gauge transformation this model is mapped onto an
Ising spin system with ferromagnetic bias; finding the ground state of
the Hamiltonian is closely related to finding the Bayes optimal
posterior under a certain noise level\cite{Sourlas}.  This mapping onto
Hamiltonian spin-systems, suggested by Sourlas for highly connected
systems, was recently extended to particular forms of sparse matrices
$A$ (where $B=I$) as well as to certain $B$ matrices~\cite{kms}. In
this extended framework, $K$ and $L$ represent the number of MSI among
the signal and noise components respectively.

Our method uses the same structure as the MN codes and builds on
insight gained from the study of physical systems with symmetric and
asymmetric\cite{ido} MSI and from examining a special case of
Gallager's method\cite{Sourlas,kms}. These theoretical
studies indicate that one may obtain superior capabilities, in terms
of the achievable code rate, by choosing high $K$ and $L$ values;
however, they come at the expense of poor decoding performance as the
corresponding basins of attraction shrink rapidly with the increasing
$K$ and $L$ values, making it essential to have high initial overlap
between the original message and the dynamical variables for the
iterative decoding process to converge successfully.  Such information
is clearly unavailable in practical scenarios. One should emphasise
that the basin of attraction shrinks dramatically. In the system
suggested by Sourlas, for instance, the initial overlap (magnetisation
in the physical system) $m=1/N \ \sum_{i=1}^{N} (2s_{i}-1)
(2s_{i}'-1)$ required in the case of $K=6$ should be higher than
$0.99$ for a successful convergence; this has been shown by numerical
simulations as well as by a mean-field calculation to be presented
elsewhere.  On the other hand, highly robust iterative decoding is
obtained for low $K$ and $L$ values at the expense of sub-optimal
capabilities (i.e., low end overlap).

The method presented here is based on constructing the matrices $A$
and $B$ in a manner that correspond to the gradual introduction of
higher connectivity sparse sub-matrices, exploiting the excellent
convergence properties of codes based on low $K$ and $L$ values with
the superior performance of high-$K$ codes.  More specifically, one
aims at starting with low MSI values, in this case $K+L \le 3$, so as
to bring the system to high overlap values from practically {\em any}
initial condition; higher values of $K$ and $L$, e.g. $3 <
K+L \le5$, may then be used for bringing the system to a perfect
overlap between the decoded and the original word.

The practical implementation of the encoding is similar to that of the
MN code except that the composed matrix used, $[A | B]$, comprises
randomly chosen sparse sub-matrices of different connectivities. The
generated codeword, constructed by taking the parity of sums of
message vector bits selected according to the specific choice of $A$
and $B$, is then transmitted through the noisy channel. Decoding the
corrupted codeword is carried out using an iterative process identical
to that of Ref.\cite{MacKay} and can take two forms: a) A gradual
introduction of higher connectivity sub-matrix components in the
Hamiltonian system used for decoding following the above description,
where end result at each stage serves as an initial condition for the
next. This is equivalent, from a physical point of view, to changing
the Hamiltonian as a function of time by gradually summing over more
message bits in Eq.(\ref{eq:hamiltonian}). b) Using the composed
matrices, including a variety of sub-matrices with different
connectivities, right from the start. The latter, which simply
correspond to a particular construction of the matrices $A$ and $B$ in
the MN code, has been used in most of our experiments due to its
simplicity, although the former has shown faster convergence at high
noise levels.  In both cases the explicit choice of sites for
generating a specific code-word bit is carried out at random, in a
similar fashion to most Gallager-type codes.

The main question that should be addressed is the optimal choice of
sub-matrix connectivities.  There are many possibilities for choosing
$K$ and $L$ values for the different stages and one should examine
various possibilities before arriving at the optimal configuration.
However, there are a few guidelines one should follow: 1) Initial
stages are characterised by low $K$ and $L$ values; $K$ values are
chosen gradually higher, so as to support the correction of faulty
bits.  2) One should choose the number of non-zero column elements as
uniformly as possible, as the number of connections per bit (spin)
defines the corrective input it receives (this is somewhat in contrast
to the approach adopted for irregular Gallager codes in which
column/row connectivity is taken from some
distribution\cite{Davey,Shokrollahi}).  3) As in most of these systems
both solutions, with $m\! =\!\pm1$, are equally attractive one should
break the inversion symmetry. This may be achieved by adding some
odd-MSI (i.e., an odd value for $K\! +\!L$) to the mainly even $K\!
+\!L$ value used initially; this assists in breaking the symmetry from
any initialisation of the iterative equations\cite{MacKay} with
practically no effect on the basin of attraction. 4) To guarantee the
inversion of the matrix $B$, and since noise bits have no explicit
correlation, we use a patterned structure, $B_{i,k} \! =\!
\delta_{i,k} \! +\! \delta_{i,k-5}$, for the $B$-submatrices with $L\!
=\!2$ and $B_{i,k} \! =\! \delta_{i,k}$ for $L\! =\!1$.  Other
practical points as well as a more detailed explanation of the
physical insight leading to the optimal choice of MSI connectivity and
the relation to Sourlas's code will be presented elsewhere.

\vspace{-0.2cm}
\begin{table}

\begin{tabular}{|c|c|c|c|c|c|c|c|c|}
R & $N$ & $A$ & $K$ & $B$ & $L$ & $f^{N}_{c}$ &
$f^{\infty}_{c}$ & $f_{c}$ \\ \hline \hline
$1/3$ & 10000 & $N\!\times\! N$ & 1 & $N\!\times\! 3N$ & 2 & 0.159 &
0.169 & 0.174 \\
  & & $3/4 \ N\!\times\! N$ & 3 & $3/4 \ N\!\times\! 3N$ & 2 & & -0.170 & \\
  & & $5/4 \ N\!\times\! N$ & 3 & $5/4 \ N\!\times\! 3N$ & 1 & & & \\
 \hline
$1/4$ & 30000 & $3/2 \ N\!\times\! N$ & 1 & $3/2 \ N\!\times\! 4N$ & 2
& 0.204 & 0.210 & 0.2145 \\
  & & $N/2 \!\times\! N$ & 3 & $N/2 \!\times\! 4N$ & 2 & & -0.211 & \\
  & & $2N \!\times\! N$ & 3 & $2N \!\times\! 4N$ & 1 & & & \\ \hline
$1/5$ & 36000 & $3N\!\times\! N$ & 1 & $3N\!\times\! 5N$ & 2 &
0.235 & 0.239 & 0.2430 \\
 &  & $2N\!\times\! N$ & 3 & $2N\!\times\! 5N$ & 1 & & -0.240 & \\ 
\end{tabular} 

\vspace*{0.1cm}
\caption{The critical flip rates $f^{N}_{c}$ and $f^{\infty}_{c}$
obtained by employing our method for various code rates in comparison
to the maximal flip rate $f_{c}$ provided by Shannon's bound. Details
of the specific architectures used and their row/column connectivities
are also provided.}
\vspace*{0.1cm}

We conclude this presentation with a demonstration of the method's
capabilities for three different code-rates $R\!=\!1/3,1/4$ and
$1/5$. In each of the cases we divided the composed matrix $[A | B]$
to six sub-matrices characterised by specific $K$ and $L$ values as
explained in table 1; the dimensionalities of the full $A$ and $B$
matrices are $M \!\times\! N$ and $M \!\times\! M$
respectively. Sub-matrix elements were chosen at random according to
the guidelines mentioned above. Encoding was carried out
straightforwardly by using the matrix $B^{-1}A$ and the corrupted
messages were decoded using the set of recursive equations of
Ref.\cite{MacKay}, using random initial conditions.  In each
case, $T$ blocks of $N$-bit unbiased messages (where exactly $1/2$ of
the bits are 1) were sent through a noisy channel of flip rate $f$
(i.e., an exact fraction $f$ of the codeword bits were flipped);
both bit and block error-rates, denoted $p_{b}$ and $p_{B}$
respectively, were monitored.  We performed at least $T\!=\!10000$
trials runs for the smaller systems ($N\!=\!10000,12000$) and
$T\!=\!1000\! -\!2000$ runs for the larger ones ($N\!=\!30000,36000$) for
each flip-rate value, starting from different initial
conditions. These were averaged to obtain the mean bit error-rate and
the corresponding variance.  In most of our experiments we observed
convergence after less than 100 iterations, except very close to the
critical flip rate.  The main halting criterion we adopted relies on
the stationarity of the first $N$ bits (i.e., the decoded message)
over a certain number of iterations. The decoding algorithm's
complexity is of $O(N)$ as all matrices are sparse. The inversion of
the matrix $B$ is carried out only once and requires $O(N^{2} \log N)$
operations.

\begin{figure}
\begin{center}
\vspace*{-1.5cm}
\epsfysize = 6.0cm
\epsfbox[0 120 670 520]{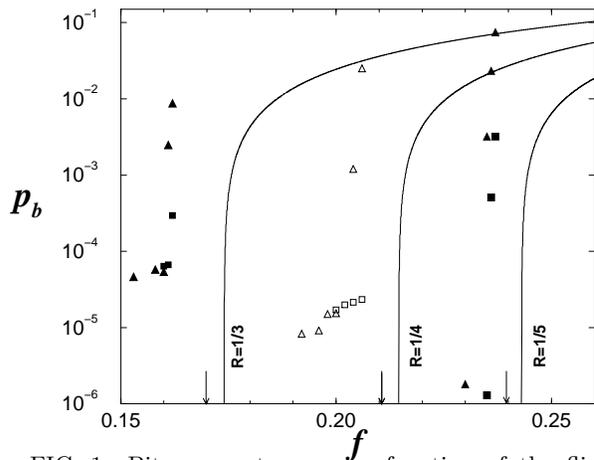}
\end{center}
\vspace*{0.5cm}
\caption{Bit-error rate $p_{b}$ as a function of the flip rate for
given code-rates $R\!=\!1/3,1/4$ and $1/5$. Results for each code-rate
appear as symbols adjacent to a line representing Shannon's
theoretical bound; triangles and squares, represent mean values
obtained for small and large network sizes respectively, corresponding
to $N\!=\!10000$ and 30000 for $R\!=\!1/3,1/4$, $N\!=\!12000$ and
36000 for $R\!=\!1/5$. Predicted code-rate values in the
$N\!\rightarrow \!\infty$ limit appear as arrows on the $x$ axis.}
\label{errors_02}
\end{figure}
 \vspace*{-3mm}

In table 1 we present the typical architectures used as well as the
maximal flip rate $f^{N}_{c}$ for which not more than a single
error-bit per block have been observed on average for a particular
message length $N$, the predicted maximal flip rate $f^{\infty}_{c}$
once finite size effects have been considered (discussed below) and
Shannon's maximal flip rate $f_{c}$ defined in
Eq.(\ref{eq:shannon_bound}).  In all these cases one obtains, on
average, perfect retrieval for noise rates that almost saturate
Shannon's bound for the critical flip rate. Just for comparison, the
corresponding results reported in Ref.\cite{Shokrollahi} for regular
and irregular Gallager codes ($R\!=\!1/4$), based on 10000 trials and
$N\!=\!16000$ report a critical value around $f\!=\!0.160$ in
comparison to $f^{N}_{c}\!=\!0.204$ and $f^{\infty}_{c}\!=\!0.210\!
-\!0.211$ reported here.

Figure 1 shows results obtained for code-rates $R\!=\!1/3, 1/4$ and
$1/5$ and various flip rates; results for each one of the code-rates
appear as symbols adjacent to a line representing Shannon's
theoretical bound for the given code-rate and noise level.  Triangles
and squares, represent mean values obtained for small and large
network sizes respectively, corresponding to $N\!=\!10000$ and 30000
for $R\!=\!1/3, 1/4$ and $N\!=\!12000$ and 36000 for $R\!=\!1/5$;
variances are smaller than the symbol size.  One notes the existence
of finite size effects, manifested in the difference between the
results obtained for different system sizes. Predicted code-rate
values in the $N\!\rightarrow \!\infty$ limit, derived below, are
represented as arrows on the $x$ axis.  The results clearly show
that in all the code-rates examined our method comes very close to
saturating Shannon's bound.

\begin{figure}
\label{errors_03}
\begin{center}
\vspace*{-1.7cm}
\epsfysize = 6.0cm
\epsfbox[0 120 670 520]{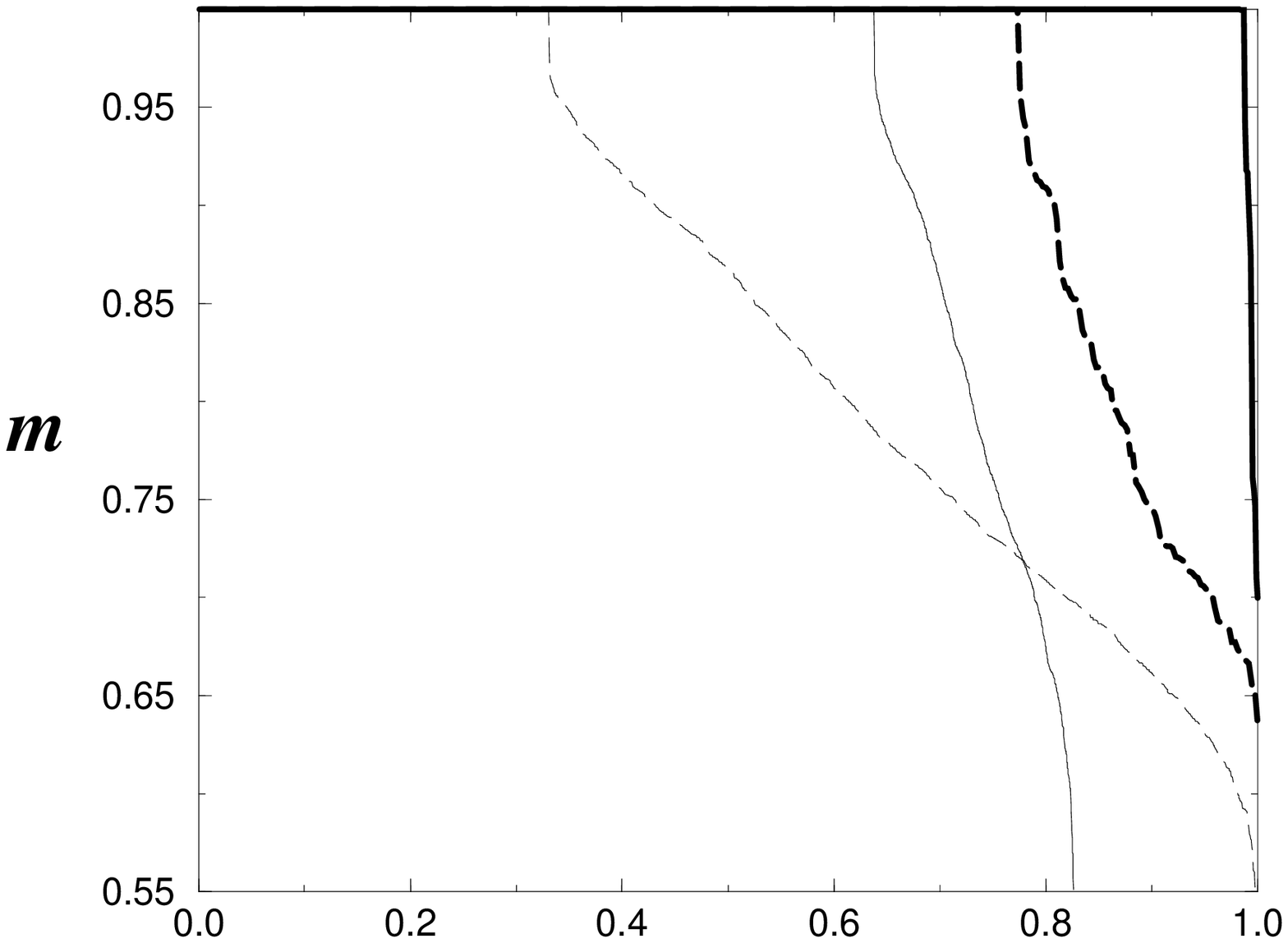}
\epsfysize = 4.7cm \epsfbox[-130 -415 420 80]{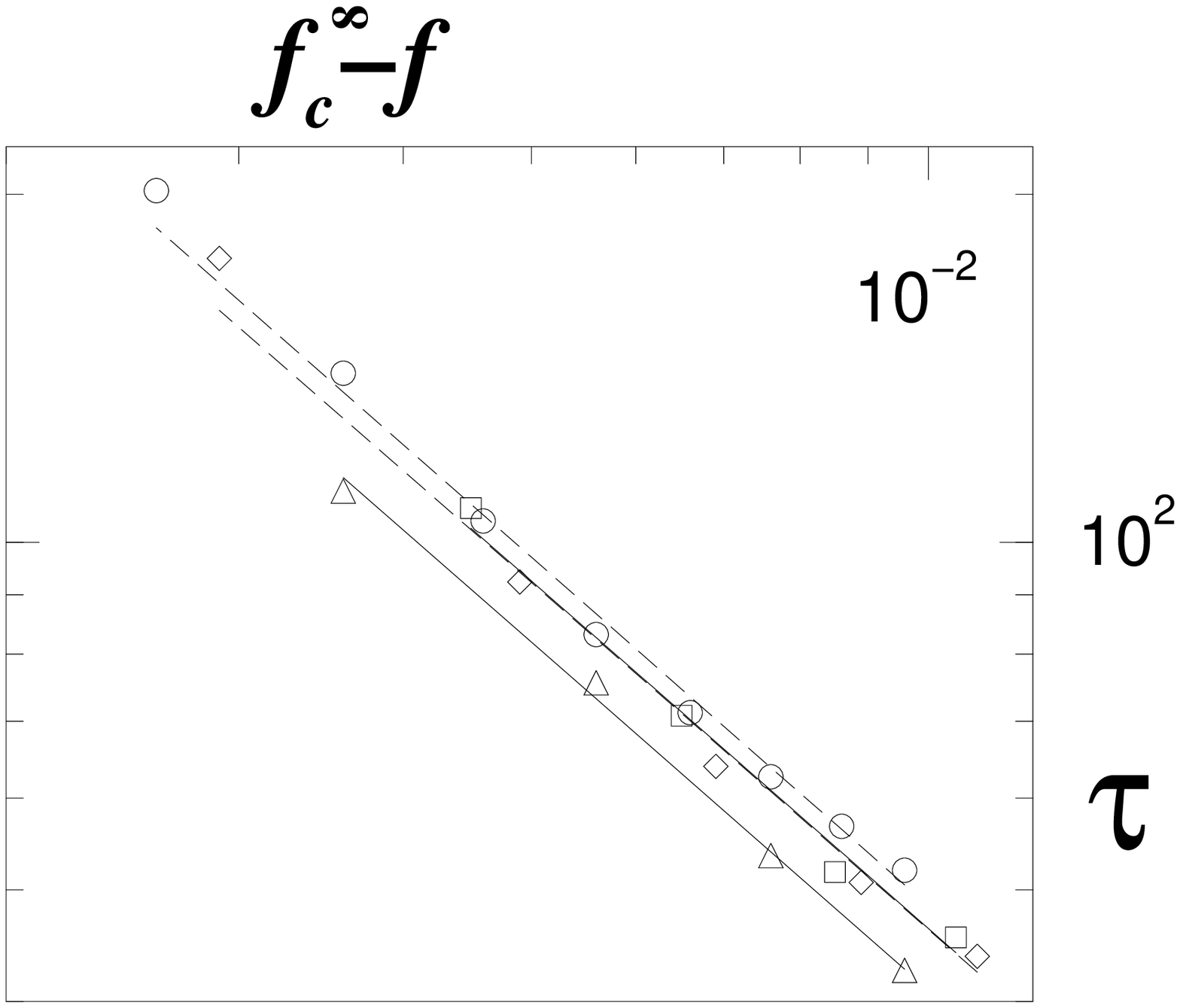}
\end{center}
\vspace*{-4.4cm}
\caption{The block magnetisations profile for $R=1/5$, $f=0.236,
0.237$ (solid and dashed lines respectively) and $N=12000,36000$,
showing the sample magnetisation $m$ vs. the fraction of the complete set
of trials. A total of 1000-10000 trials (for larger and smaller
systems respectively) were rearranged in a descending order according
to their magnetisation values (directly related to the overlap between
the decoded and the original message). The fraction
of perfectly retrieved blocks increases with system size (thick
lines). Inset - log-log plots of mean convergence times $\tau$ for
$R=1/3$ and $N=10000$ ($\bigtriangleup$), $R=1/4$, $N=10000$ ($\Box$)
and $N=30000$ ({\large $\diamond$}), $R=1/5$ and $N=36000$ ({\large
$\circ$}). The $f^{\infty}_{c}$ values were calculated by fitting
expressions of the form $\tau \propto 1/(f^{\infty}_{c}-f)$ through
the data (dashed lines for the larger systems).}
\end{figure}
 \vspace*{-3mm}

The results shown so far are based on finite-$N$ simulation results.
However, as Shannon's bound itself is based on infinitely large
messages, one cannot expect to saturate the bound completely for
finite-$N$ messages.  To assess the critical flip rate achievable by
our method in the limit of infinitely large systems, $f^{\infty}_{c}$,
we monitor two criticality indicators: a) The dependence of the
block error distribution on the system size - the transition from
perfect($p_{B}(f)\!=\!1$) to no retrieval ($p_{B}(f)\!=\!0$), as a
function of the flip-rate $f$, is expected to become a step function (at
$f^{\infty}_{c}$) as $N\!\rightarrow\! \infty$. If the percentage of
perfectly retrieved blocks in the sample, for a given flip rate $f$,
increases (decreases) with $N$ one can deduce that $f\! <\!
f^{\infty}_{c}$ (or $f\! >\! f^{\infty}_{c}$). b) Convergence times as a
function of $f$ - convergence times near criticality usually diverge
as $1/(f^{\infty}_{c}-f)$; by monitoring average convergence times for
various $f$ values and extrapolating one may deduce the corresponding
critical flip rate.

In Fig.2 we ordered the samples obtained for $R\!=\!1/5$,
$f\!=\!0.236, 0.237$ (solid and dashed lines respectively) and
$N\!=\!12000,36000$ according to their magnetisation; results with
higher magnetisation appear on the left and the $x$ axis was
normalised to represent fractions of the complete set of trials. One
can easily see that the fraction of perfectly retrieved blocks
increases with system size (thick lines) indicating that $f <
f^{\infty}_{c}$. Repeating the same exercise for higher $f$ values we
obtained an estimate of $f^{\infty}_{c}$ reported in table 1. In the
inset one finds log-log plots of the mean convergence times $\tau$
for $R\!=\!1/3,1/4,1/5$ and different $N$ values, carried out on
perfectly retrieved blocks with less than 2 error bits. The optimal
fitting of expressions of the form $\tau \propto 1/(f^{\infty}_{c}-f)$
through the data provides another indication for the $f^{\infty}_{c}$
values, which are consistent with those obtained by the first method.

To conclude, we have shown that through a successive change in MSI and
connectivity, while keeping the connectivity low ($\le 5$), one can
boost the performance of matrix based error-correcting codes, getting
ever closer to saturating the theoretical bounds set by Shannon.  It
is quite plausible that the performance reported here may be improved
upon by fine tuning the choice of architecture, which is currently
under way. Moreover, it is highly likely that several architectures
will provide similar performance in the thermodynamic limit; it would
be worthwhile to examine their finite size behaviour above and below
saturation which is of great practical significance.

\end{table}


\vspace{-0.5cm}
\begin{thebibliography}{99}
\vspace{-1.7cm}
\bibitem{Shannon} C.E.~Shannon, {\em Bell Sys.Tech.J.}, {\bf 27}, 379
 and 623 (1948).
%
\bibitem{Gallager} R.G.~Gallager,  {\em Low density parity check codes}
Research monograph series {\bf 21} (MIT press), 1963.
%
\bibitem{Cover} T.M.~Cover and J.A.~Thomas, {\em Elements of
 Information Theory} (Wiley), 1991.
%
\bibitem{err_cor_book} A.M.~Michelson and A.H.~Levesque,
{\em Error-Control Techniques for Digital Communications} (Wiley), 1985.
%
\bibitem{turbo}
C.~Berrou and
A.~Glavieux {\em IEEE Trans.Comm.}, {\bf 44}, 1261 (1996).
%
\bibitem{MacKay}
D.J.C.~MacKay, {\em IEEE Trans. IT}, {\bf 45},
399 (1999).
%
\bibitem{Davey} M.C.~Davey  and D.J.C.~MacKay, {\em IEEE Comm. Lett.},
in press (1999).
%
\bibitem{Shokrollahi} M.~Luby,  M.~Mitzenmacher,  A.~Shokrollahi and
D.~Spielman, in {\em IEEE proceedings of the International
Symposium on Information Theory ISIT98} (1998).
%
\bibitem{Frey} B.J.~Frey, {\em Graphical Models for Machine Learning
and Digital Communication} (MIT Press), 1998.
%
\bibitem{Sourlas} N.~Sourlas,  {\em Nature}, {\bf 339} 693 (1989).
%
\bibitem{kms} Y.~Kabashima and D.~Saad, {\em Euro.Phys.Lett.}, {\bf
45} 97 (1999); Y.~Kabashima, T.~Murayama and D.~Saad, unpublished
(1999).
%
\bibitem{ido} I.~Kanter {\em Phys.Rev. A}, {\bf 38,} 5972 (1988), and
I.~Kanter and H.~Sompolinsky, {\em Phys.Rev.Lett.}  {\bf 58}, 164
(1987).

\end{thebibliography}
\end{document}